# Scattering phases in the broken phase of the 4-d $O(4)$ non-linear $\sigma$-model

Jörg Westphalen[a,b*], Frank Zimmermann[a,b],
Meinulf Göckeler[a,b†] and Hans A. Kastrup[a]

[a]Institut für Theoretische Physik E, RWTH Aachen, D-52056 Aachen, Germany

[b]HLRZ c/o KFA Jülich, P.O.Box 1913, D-52425 Jülich, Germany

Using Lüscher's method we determine the elastic scattering phases in the broken phase of the 4-dimensional $O(4)$ non-linear $\sigma$-model from the two-particle energy spectrum in a Monte-Carlo study on finite lattices. In the isospin-0-channel we observe the $\sigma$-resonance and extract its mass and its width. In all scattering channels investigated the results are consistent with perturbative calculations.

## 1. INTRODUCTION

Lüscher established a relation between the energy spectrum of two-particle states in a finite box with periodic boundary conditions and elastic scattering phase shifts defined in infinite volume [1]. Since two-particle energy levels are calculable by Monte Carlo techniques, this relation opens the possibility to extract phase shifts from numerical simulations on finite lattices. We give a résumé of our results on the elastic scattering phases in the broken phase of the 4-dimensional $O(4)$ non-linear $\sigma$-model including the corrections to our preliminary results of the Lattice '92 contribution [2]. For a more detailed discussion see ref. [3].

## 2. FROM THE TWO-PARTICLE SPECTRUM TO THE SCATTERING PHASES

We give the "pions" of our model a nonzero mass $m_\pi$ by means of an external source $J$ in the action:

$$S = -2\kappa \sum_x \sum_{\mu=1}^4 \Phi_x^\alpha \Phi_{x+\hat\mu}^\alpha + J \sum_x \Phi_x^4 . \quad (1)$$

The scalar field is represented as a four-component vector $\Phi_x^\alpha$ of unit length: $\Phi_x^\alpha \Phi_x^\alpha = 1$.

*speaker at the conference
†supported by the *Deutsche Forschungsgemeinschaft*

Two-pion states are classified according to cubic symmetry and have "isospin" $I = 0, 1, 2$. For their investigation we define the operator:

$$\mathcal{O}_i^{ab}(t) = \sum_{\vec n \in \mathbb{Z}_L^3} \tilde f_i(\vec n)\, \tilde\Phi_{-\vec n,t}^a\, \tilde\Phi_{\vec n,t}^b \quad (2)$$

where $\tilde f_i(\vec n)$, $i = 1, 2, \ldots$ is some basis of wave functions with correct cubic symmetry and $\tilde\Phi_{\vec n,t}^a$ is the spatial Fourier transform of the field $\Phi_{\vec x,t}^a$ on a lattice of spatial extent $L$:

$$\tilde\Phi_{\vec n,t}^a = L^{-3} \sum_{\vec x \in \mathbb{Z}_L^3} \Phi_{\vec x,t}^a\, e^{2\pi i \vec x \cdot \vec n / L} . \quad (3)$$

The simplest choice of cubically invariant wave functions is a sum over plane waves:

$$\tilde f_i(\vec n) = \delta_{i,\vec n^2} , \quad i = 0, 1, 2, \ldots . \quad (4)$$

We also used another set of (Lüscher−) wave functions [1], which gave the same final results (see [3] for further details).

In the isospin-0 channel, where the $\sigma$-resonance is expected, we use operators $O_i(t)$ given by

$$O_i(t) = \frac{1}{\sqrt{3}} \sum_{a=1}^3 \mathcal{O}_i^{aa}(t) . \quad (5)$$

Additionally we have to take into account the $\sigma$ field at zero momentum

$$O_\sigma(t) = \tilde\Phi_{\vec 0,t}^4 = \frac{1}{L^3} \sum_{\vec x \in \mathbb{Z}_L^3} \Phi_{\vec x,t}^4 , \quad (6)$$

since it has the correct quantum numbers and is expected to create a state with energy below the inelastic threshold.

The two-particle energies $W_\nu$ are extracted from the matrices of connected correlation functions [4]. For the different isospin channels these matrices are given by

$$C_{ij}^0(t) = \langle O_i(t) O_j(0) \rangle_c,$$
$$C_{ij}^1(t) = \langle \operatorname{Im} \mathcal{O}_i^{ab}(t) \operatorname{Im} \mathcal{O}_j^{ab}(0) \rangle_c, \quad (7)$$
$$C_{ij}^2(t) = \langle \operatorname{Re} \mathcal{O}_i^{ab}(t) \operatorname{Re} \mathcal{O}_j^{ab}(0) \rangle_c - C_{ij}^0(t).$$

As described in [1], for each lattice extent $L$ and each two-particle energy level $W_\nu < 4m_\pi$ we get one value of the scattering phase shifts $\delta_0^0$, $\delta_1^1$, and $\delta_0^2$, respectively, in the elastic region $0 < k_\nu / m_\pi < \sqrt{3}$. It is computed from the key relation

$$\delta_l^I(k_\nu) = -\phi\left(\frac{k_\nu L}{2\pi}\right) \mod \pi, \quad (8)$$

where $\phi$ is a continuous function defined in [1], see also [2,3] ($l$=angular momentum). The momentum $k_\nu$ corresponding to $W_\nu$ has to be calculated with the help of the (lattice) energy momentum relation

$$\left(2 \sinh \frac{(W_\nu/2)}{2}\right)^2 = m_\pi^2 + \vec{\tilde{k}}_\nu{}^2. \quad (9)$$

This relation does not determine $k_\nu$ uniquely, since $\vec{\tilde{k}}_\nu{}^2$ depends not only on $k_\nu$ but also on the direction of $\vec{k}_\nu$. We include this small ambiguity in the error estimate for $k_\nu$.

## 3. NUMERICAL RESULTS

The data points in figs. 1- 3 show the momenta extracted from the energy levels and the corresponding phase shifts $\delta_0^0, \delta_1^1, \delta_0^2$ at ($\kappa = 0.315$, $J = 0.01$).

How do our results compare with one–loop perturbation theory if we insert the renormalized coupling constant and masses as determined by our simulations ? The dashed curves are the perturbative predictions based on an estimate $\tilde{m}_\sigma$ of the resonance mass $m_\sigma$, which was obtained from a fit to the non–resonance $\sigma$-propagator in momentum space. We consider $\tilde{m}_\sigma$ to be an estimate only, since the $\sigma$ particle is unstable. Fig. 1 shows that indeed $m_\sigma$ lies below $\tilde{m}_\sigma$.

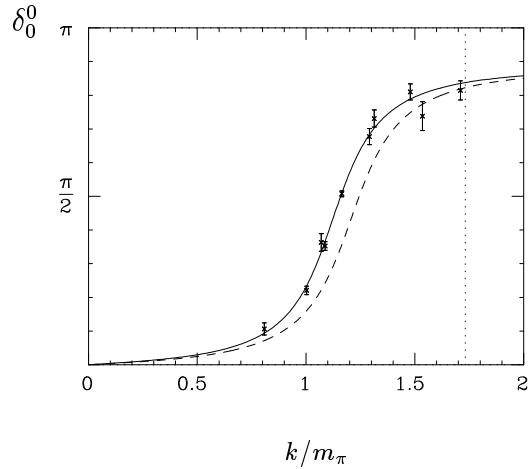

Figure 1. Comparison of the perturbative fit 3.A. (solid curve) and the perturbative prediction (dashed curve) in the isospin-0 channel.

In figs. 2, 3 we observe no significant deviation from perturbation theory: For isospin 1 and 2 the perturbative predictions depend on $m_\sigma$ only weakly.

In order to determine the resonance mass $m_\sigma$ and the decay width $\Gamma_\sigma$ from the measured scattering phases we have employed different methods:

### A. Perturbative Fit

The one-loop perturbative formula for $\delta_0^0$ [5] (see [3] for details) as function of $k; m_\pi, m_\sigma, \Gamma_\sigma$ can be used as a fit ansatz for *all* points in the elastic region. This leads to $m_\sigma = 0.691(3)$ and $\Gamma_\sigma = 0.112(7)$ in good agreement with the results from the Breit-Wigner fit below.

### B. Breit-Wigner Fit

Fitting to the (relativistic) Breit-Wigner formula

$$\tan\left(\delta_0^0 - \frac{\pi}{2}\right) = \frac{W^2 - m_\sigma^2}{m_\sigma \Gamma_\sigma} \quad (10)$$

has the major advantage of being free of additional asumptions, but can only be applied near the resonance at $\delta_0^0 = \pi/2$, so only few data points are used. Nevertheless we get for the resonance mass $m_\sigma$ and width $\Gamma_\sigma$: $m_\sigma = 0.706(2)$ and $\Gamma_\sigma = 0.130(9)$. These numbers have to be compared with the estimate $\tilde{m}_\sigma = 0.720(1)$ and the perturbative prediction $\tilde{\Gamma}_\sigma = \Gamma_\sigma(\tilde{m}_\sigma, \ldots) = 0.121(1)$.

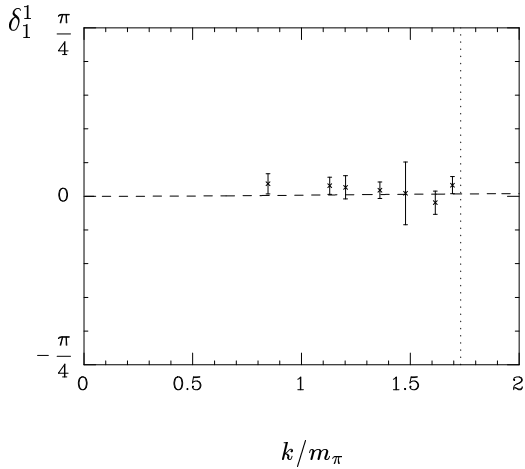

Figure 2. *Is there a $\rho$ in the isospin-1 channel? – The scattering phase shifts for isospin 1 indicate that there is none as expected from perturbation theory.*

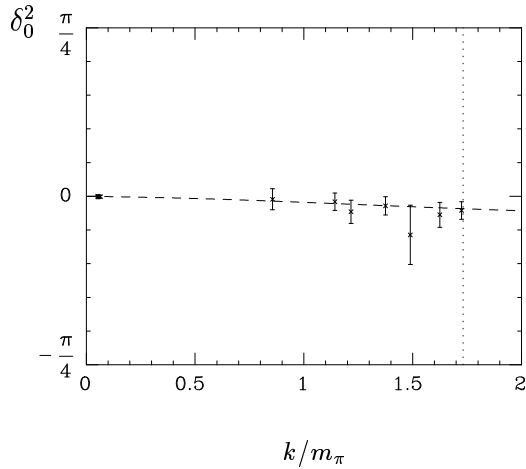

Figure 3. *Scattering phase shifts for isospin 2.*

## 4. DISCUSSION AND CONCLUSIONS

The results for the decay width agree reasonably well with the perturbative predictions, while the resonance masses $m_\sigma$ lie systematically (about 5 %) below the estimates $\tilde{m}_\sigma$ from the fit to the propagator in momentum space. However, this discrepancy should not be too surprising, because the width of the $\sigma$-resonance for our choice of parameters is rather large: $\Gamma_\sigma \approx 0.15\, m_\sigma$.

Once again, renormalized perturbation theory has turned out to be very reliable in the four-dimensional $\Phi^4$-theory. Furthermore, we have demonstrated the applicability of Lüscher's method for studying particle scattering processes in massive quantum field theories on finite lattices – at least for this model.

## ACKNOWLEDGEMENT

Helpful discussions with M. Lüscher and C. Frick are gratefully acknowledged. Furthermore we wish to thank the *Rechenzentrum* at the RWTH Aachen and the HLRZ Jülich for providing the necessary computer time on their SNI S600/20 and CRAY Y-MP, respectively.